# Reckoning with the wicked problems of nuclear technology: Philosophy, design, and pedagogical method underlying a course on Nuclear Technology, Policy, and Society

Aditi Verma[1]
Assistant Professor of Nuclear Engineering and Radiological Sciences
University of Michigan

## Introduction

With a growing global emphasis on the need to decarbonize energy systems, key decision-makers in many countries are calling for a significant expansion of nuclear energy – with projections calling for a doubling or tripling of nuclear capacity around the globe by mid-century.[1] These expansions in capacity projected, include in many cases, the use of nuclear technologies in entirely new places and contexts – countries that have never before built and operated nuclear reactors, applications of nuclear energy technologies in remote and off-the-grid locations, as well as the use of nuclear technologies for generating process heat for a wide variety of industrial applications. Beyond the technical work needed to rapidly develop and commercialize nuclear energy systems over these aggressive timescales for use across a potentially vast set of use contexts, a range of problems – political, ethical, social, environmental, and economic – must also be acknowledged and explored if nuclear energy systems are to become significantly integrated into our energy systems of the future.[2] Doing so requires that nuclear engineers of the future must be equipped with the intellectual frameworks and tools to make sense of these problems – both as they manifest in the context of nuclear technologies and industries as we know them today and learn to anticipate the forms these problems might take in the future. Traditionally, these skills and ways of thinking often consigned to the category of "non-technical" or "soft", have been relegated to footnotes and sidebar discussions, never occupying a central place in the intellectual canon of nuclear engineering. Yet a growing number of young people [3,4] entering the discipline of nuclear engineering, are increasingly expressing an interest in reckoning with these difficult – or what we will call here, the "wicked problems" of nuclear engineering.

Wicked problems — a formulation put forward in what is now a landmark paper by Rittel and Webber [5] (design and planning scholars respectively) – are those that lack definitive formulations, resist durable resolution, do not have an exhaustively identifiable set of true or false solutions, and are often framed entirely differently by different entities experiencing the problem. Every attempt to solve a wicked problem is a solution attempt made in the real world and thus has consequences and implications that can potentially be far-reaching. (A fuller discussion of wicked problems in a nuclear context is contained in the subsequent section. See also Table 1).

Many, if not all problems, at the intersection of nuclear technology, policy, and society bear all these (and other, as shown in Table 1 below) hallmarks of wickedness. These include difficulties within the nuclear sector in seeing through new technological designs from the inception to implementation[6]; cost overruns, financing difficulties, and the recent mismanagement of nuclear plant construction projects[7]; the presence of regulatory institutional infrastructures that in both real and perceived waus can curtail meaningful learning; a rigid way of thinking about risk and safety within the nuclear sector which have unintentionally led to the creation of an antagonistic expert-public divide[2]; our failures to successfully engage communities in the siting and technology development process [8]; the still unresolved problem of long-term nuclear waste management which even if regarded as a technically solved problem, remains unresolved in a real sense[9]; the dual-use nature of nuclear technologies that create security and non-proliferation concerns; and a myriad of environmental justice issues that pervade the nuclear fuel cycle and even several aspects of nuclear policymaking. [10] The list is not exhaustive. It goes on and is subject to change. The form these problems take is likely to change as our technologies evolve and as they are potentially used on an ever larger scale – as we intend them to be. What then is our responsibility to society? Or more immediately, what is our responsibility to future nuclear engineers? How must we prepare them to reckon with these problems?

To these questions, this paper offers some answers. This paper describes the underlying philosophy, design, and implementation of a course on "Nuclear Technology, Policy, and Society" taught in the Department of Nuclear  Engineering and Radiological Sciences at the University of Michigan. The course explores some of nuclear technology's most pressing challenges (or its 'wicked problems'). Through this course students explore the origins of these problems – be they social or technical, they are offered tools – conceptual and methodological –  to make sense of these problems, and guided through a semester-long exploration of how engineers can work towards their resolution, and to what degree these problems can be solved through institutional transformation and/or a transformation in our own practices and norms as a field. The underlying pedagogical philosophy, implementation, and response to the course are described here for other instructors who might wish to create a similar course, or for non-academic nuclear engineers, who might perhaps, in these pages, find a vocabulary for articulating and reflecting on the nature of these problems as encountered in their praxis. The  paper is structured as follows: a background section describes the emergence of ethics as an area of inquiry and education as well as prior policy-related pedagogical efforts at the University of Michigan. This is followed by a brief section on the contributions of this paper and the course that it describes, followed by a section on method which describes the course philosophy and design. This section in turn is followed by a discussion of course outcomes, recommendations to others considering developing or teaching similar courses, and a section that concludes.

**Background**
With the creation of the first engineering ethics standards in the late nineteenth century, many engineering societies began to draft their own codes of ethics by the early twentieth century. [11] In the nuclear sector – this has included codes of ethics for both academic and professional engineers (as created by the American Nuclear Society in the US[12], for example) as well as codes of ethics for nuclear operating organizations (as created by the International Atomic Energy Agency[13]). With the widespread adoption of codes of ethics, the Accreditation Board for Engineering and Technology (ABET) also started calling for the inclusion of ethics in engineering education and began auditing its presence as part of its accreditation processes. Yet, what ethics means, and how it is implemented in engineering pedagogy, varies widely across engineering disciplines – as does the understanding of what constitutes the ethical responsibility of an engineer[14,15]. This, in part, has to do with the breadth of expertise of engineering instructors, their ability and interest in teaching ethics, as well as institutional barriers and constraints that might prevent the meaningful inclusion of ethics, as well as other social and policy considerations in engineering education[16,17]. Many have critiqued the concept of ethics itself as operationalized in engineering education as being too narrow, calling instead for a focus not just on ethics but on justice. [18,19] Reviews of engineering literature reveal that until recently engineers have typically been hesitant to adopt a pro-justice positionality[20]. One possible explanation of this is that the adoption of such a stance is an implicit acceptance of the political nature of engineering – an ideology to which engineers have long objected. Yet a growing body of research on how engineers and designers make design decisions, how they often unintentionally encode their designs with their values and biases, designing for a 'reference' or individual user they are able to easily imagine, while often failing to recognize the needs of many others, is leading to a growing consensus across fields of engineering that engineers need to frame and solve design problems keeping in mind a much broader range of possible users and constituents, and design with these individuals and communities wherever possible. [21,22] These logics which are rapidly becoming mainstream in fields such as product design [23], AI[24], and robotics, [25] are gradually also reaching fields of engineering concerned with the design of complex systems – systems that do not have a single or even a handful of users but instead a complex web of rightsholders.

Interdisciplinary Nuclear Technology Studies at the University of Michigan

It is significant that the course described here is being taught at the University of Michigan – home to the first research initiative dedicated to the interdisciplinary study of nuclear energy. This initiative — the Michigan Memorial Phoenix Project was created in 1948 – is headquartered in a physical building on the University of Michigan's North Campus at 2301 Bonisteel Boulevard. The Memorial was imagined as " a living memorial to the 585 university alumni, students, faculty, and staff members who gave their lives in World War II" devoted to the "peaceful, useful, and beneficial applications and implications of nuclear science and technology for the welfare of the human race".[26] The Memorial, which predated the Department of Nuclear Engineering and Radiological Sciences, brought together researchers of many intellectual stripes – physicists, engineers, lawyers, political scientists, and sociologists, to study nuclear energy and its applications. Regrettably, over time, this momentum faded, and the purpose and mission of the memorial were all but forgotten until 2022 when the Memorial was rededicated to its original mission (an initiative led by the new department chair, Professor Todd Allen). The preceding year also marked the return of the Department of Nuclear Engineering and Radiological Sciences to the Michigan Memorial Phoenix Project building. The Phoenix building is also currently home to Fastest Path to Zero – an interdisciplinary initiative at the University of Michigan whose purpose is to help communities reach their climate goals.

A brief history of curricular development at the University of Michigan

Even preceding these developments, various faculty at the University of Michigan have taught courses on the history and policy of nuclear technology. Most recently, Professor Todd Allen taught a course on nuclear policy. The course, taught in a hybrid fashion in Fall 2020 (during the pandemic), brought a diverse array of policymakers who lectured on the structuring and functioning of their respective institutions. Dr. Patricia Schuster taught a course on the history of nuclear weapons. This course was offered in Fall 2018. Going further back, NERS Professor Emeritus Ron Fleming briefly taught a course on the history of nuclear weapons, and Professor Gabrielle Hecht, during her tenure in the history department, taught courses on the history of nuclear energy. All but Hecht's courses in the History department were one-off offerings, and Hecht's move to Stanford University created an even larger vacuum. Following widespread student demand (including an initiative by some students to themselves create and teach a course – an initiative that ultimately was not realized), and on the recommendations of the Department's Advisory Board (a group of senior practitioners, researchers, and policymakers), a decision was made to create a new, recurring course – which is the course created by the author and described in this paper.

*The instructor's background*

An ideology of logical positivism and depoliticization[27] pervades the natural sciences and engineering. Peer-reviewed writings are written in the third person, describing the logical order of research questions, methods, results, analysis, and discussion – thus projecting a veneer of pure objectivity and linearity in the research process. Often, the actual research process could not be further from reality. Bold discoveries and inventions are sometimes purely accidental – not the result of deliberate planning and design, but fortunate serendipity,[28–30] and often unexpected results that precede the research questions, with the results once realized prompting researchers to return to an earlier phase of exploration and self-inquiry to ponder whether the correct questions were being asked or indeed what are the questions to which the unexpected results are the answer. Observing the circuitous nature of the process of research and discovery but the nevertheless linear account of it presented in publications has prompted many researchers, including a Peter Medawar – a Nobel Laureate in Medicine (1960), to speculate whether the structure of a research paper is a 'fraud'.[31] The structure of the research paper tells us little about the manner in which the actual work proceeded and indeed effaces all the quirks and happenstances that are a feature, not a bug, of the research process. Drawing from another body of work – the study of design – one might go even further to say that yet another important aspect of research the research process is obscured – the identity and expertise of the researcher/author, their background, and how their particular set of proclivities led them to the results described in the paper.[32,33] In an emerging tradition in many fields of study, authors start by describing their own backgrounds and why understanding their backgrounds is essential to understanding their work. It is with this emerging tradition that I align myself. Though the remainder of the paper is written in the third person, I pause here to explain my own background as a way for the reader to make sense of the key decisions made in the design and implementation of this course.

I was drawn as a high school student to the field of nuclear engineering precisely because of its policy problems. Having grown up in India where, even during the 1990s, blackouts, and brownouts were a frequent occurrence, I viewed nuclear energy technologies as having enormous potential for good. A handful of experiences participating in simulated sessions of the IAEA helped me realize that nuclear technology's policy problems could only be solved by having a deep understanding of the science and engineering underlying the technology as well as access to conceptual frameworks and tools that lay outside the field – in history, economics, political science, and many other disciplines.

Though initially intending to major in both physics and economics, I ultimately chose nuclear engineering. The choice of a signal major made it possible, once I had completed my core coursework as an undergraduate and then doctoral student, to take many elective courses from other departments. This interdisciplinary coursework included courses on theories of the state and economy; theories of innovation; comparative political economy; energy economics; engineering, regulation, and management of the electric power sector; design of social science research projects; and qualitative research methods.

This coursework was supplemented with extensive reading (both supervised and unsupervised) on foundational texts in social theory, the history, and sociology of technology, design research, and a broad (though not exhaustive as the field is immense) selection of works from risk studies. The broad range of conceptual frameworks and methodological tools drawn from these courses and readings have shaped key decisions in the design of this particular course. For example, as described in the section on underlying course philosophy, I explain how in every course session that focuses on a particular policy problem, students learn about the empirical status and framing of that problem, while also learning to view that problem through one or more conceptual frameworks – many of which were gleaned from the coursework and reading described above and which continue to shape my thinking today. My growing interdisciplinary intellectual identity took further shape in my thesis project – a study of how reactor designers make decisions in the early, foundation stages of design. This was an effort that brought to bear my grounding in nuclear engineering with research methods from the social sciences, and theoretical frameworks from the field of design research. Each of these disciplines was represented on my dissertation committee.

The decade of study and research at the Massachusetts Institute of Technology was followed by a two-year position at the OECD Nuclear Energy Agency where a final tool in my toolbox was acquired – an understanding of how policymakers conceptualize and make decisions about the problems that had drawn me to the field. This experience too shapes the course. As described in the section on underlying philosophy – a key tenet of the course is the need for students to be able to understand problems at the intersection of nuclear technology policy, and society from multiple, even conflicting, perspectives.

A return to academia at the Harvard Kennedy School of Government for a postdoctoral fellowship led to a new area of inquiry – the study of composition and pedagogical practices, particularly as they pertain to design across engineering as well as non-engineering disciplines. How did designers – engineers or not – engage (or not) with society and how? How did they make sense of their responsibilities? A review of over 200 syllabi yielded important insights – that few courses treated questions of ethics and justice meaningfully, and many that did tend to over-intellectualize these questions –removing from the lived experience of people, from reality [14]. These learnings too shape this course. Writings by community members, even activists, appear alongside readings by researchers and policymakers. As part of course assignments, students engage with people to explore their understandings of and views on nuclear technology.

**Contributions and Underlying Course Philosophy**

The contributions of the course are best understood by unpacking how the course prepared students to rigorously approach problems at the intersection of technology, policy, and society.

A central tenet of the course is that students must learn not only what to think but how to think about a broad range of problems at the intersection of nuclear technology, policy, and society. This is because problems framed in a certain way today might be framed entirely differently in the future and further, problems regarded as pressing today may not be as pressing in the future, with still other problems, we have not yet imagined, rising to the fore. For example, a relatively well-developed system of export controls and safeguards is applied today particularly to countries that are "non-nuclear" weapons states and make use of nuclear technologies for peaceful applications. However, in a potential future when no new nuclear weapons remain, instruments such as safeguards and even export controls may need to be applied even more rigorously to prevent re-armament or the re-development of nuclear weapons. Another example might be drawn from nuclear waste management – while waste volumes are relatively small today and require the identification of one or at most a couple of sites for deep geological repositories in each country using nuclear energy technologies, a future in which nuclear energy technologies – both fission and fusion – are potentially used at a much larger scale, is one in which waste volumes might be sufficiently significant to prompt concern and require siting of multiple, possibly even highly localized repositories. As a third example, though uranium resources are regarded as plentiful today, a possible expansion of fission energy technologies might prompt a surge in demand, leading either to an increase in mining (which would require weighing the benefits of procuring uranium against the environmental and social harms of mining) or a normalization of reprocessing technologies. When the time comes, and as the tenor and framing of nuclear policy problems change, nuclear professionals of tomorrow – our students today – need to be prepared to knowledgeably grapple with these problems, however implausible the problems might seem today.

Just as students learn analytical rigor in their other engineering courses, so too must analytical rigor be applied to problems at the intersection of technology, policy, and society, even if these problems seldom, if ever, yield to everlasting solutions. In this course students learn the ability to approach these problems rigorously by:

(1) Understanding multiple, even conflicting perspectives on the same problems. For example in an early course session on the role of nuclear energy in decarbonized energy systems, students learn arguments put forward both for and against nuclear energy technologies and learn how to weigh these arguments without vilifying the originators of the arguments.

(2) Understanding multiple framings of an identical problem from numerous standpoints. Students examine nuclear policy problems as framed by engineers, citizens and community members, and policymakers themselves. For example, in a session about the siting of new nuclear energy facilities, students learn the state and federal approaches to siting facilities, they read research studies by scholars from the field of planning, while also reading perspectives from community-based organizations offering support and critique of said facility. The goal is not to single out a particular framing of the problem as right or wrong – but rather to make sense of why so many perspectives are able to co-exist and what the origins of diverging framings might be, including how the positionality of the persons or organizations approaching the problem, shapes their framing of it.

(3) Linking theory to practice. Students are offered one or more conceptual and theoretical frameworks through which to make sense of the policy problems they are examining. For example, initial course sessions focus on the history of nuclear energy technology development and the emergence of light water reactors as the leading or dominant technology. To make sense of this history – students complete and are led through a discussion of research from the sociology of technology on the social construction of technology[34], interpretive flexibility, and technological momentum[35] and lock-in[36] alongside research on theories of innovation on creative destruction[37], disruptive innovation[38], and the emergence of dominant designs[39] as well as critiques of innovation itself.[40,41] Similarly, in a following session on the global nature of the nuclear industry and international transfers of technology [42,43], students learn about theories about early and late industrial development[44] and technological leapfrogging[45,46], alongside work on complex product systems[47] – of which nuclear reactors are an archetypal example. While it is true that entire courses could be offered on these theories (and are) in political science and economics departments, the intent behind including these theories and conceptual frameworks in an applied manner in the course is to show students how practice and empirical reality can be related to theory; how other disciplines, including seemingly intellectually distant ones, might offer crucial tools for grappling with problems at the intersection of technology policy, and society; to imbibe enough of the language and jargon from these disciplines to be able to immerse themselves further and engage with scholars and practitioners from these disciplines, if they so desire; and most of all to understand the limits of their own expertise and appreciate that of others, while being willing to transgress traditional disciplinary boundaries in service of a greater good, as needed.

**Method: Course design**

This section of the paper describes the underlying course philosophy as well as course design.

Framing nuclear technology problems as 'wicked problems'

The course begins by framing problems at the interaction of technology, policy, and society as wicked problems. Table 1 below shows the ten attributes of wicked problems as laid out by Rittel and Webber.[5] It also describes how 'tame' problems differ from wicked problems and offers examples of nuclear problems as 'wicked problems.' In the first course session, having learned about wicked problems and their attributes, students are invited to reflect on whether problems at the intersection of nuclear technology, policy, and society, of which they are aware, exhibit these different attributes of wickedness. (This initial course session is followed by a related assignment described below.) The wicked problem framing has proved to be very useful in the course as we often return to this way of thinking about problems at the intersection of technology, policy, and society several times during the semester including during sessions on nuclear waste, safety, innovation, siting of new nuclear facilities, economics, and proliferation.

**Table 1.** This table describes the 10 attributes of wicked problems as laid out by Rittel and Webber, with brief examples of nuclear technology problems that illustrate their 'wicked' nature.

| Attributes of wicked problems[5] | How 'tame' problems differ from wicked problems (paraphrased from Rittel and Webber)[5] | Further description of the wicked problem attribute(paraphrased from Rittel and Webber)[5] | A nuclear example |
|---|---|---|---|
| 1. Do not have a definitive formulation | "tame problems" have exhaustive formulations containing all the information needed to solve the problem. | "The information needed to understand a problem depends on one's idea for solving it." The understanding and solution of the problem are inextricably linked to each other "One cannot understand the problem without knowing about its context; one cannot meaningfully search for information without the orientation of a solution concept" | <u>Nuclear waste</u> Is spent fuel waste? Should it be buried or recycled? Should we build in retrievability? |
| 2. Do not have a stopping rule | The problem solver knows when they have done their job. There are criteria that indicate or specify that a solution has been met. | "No ends to the causal chains that link interacting open systems". The planner can always do better. Additional investment of effort might lead to a better solution. A planner stops working on a problem not because it is perfectly solved but because the planner has run out of time. The planner has to satisfice. | <u>Nuclear safety</u> We will always need to remain vigilant about nuclear safety, always need to ensure we have public consent, and always need to ensure that nuclear technologies and materials are not used to build weapons. Nuclear regulators can always do additional things to make plants safer. |
| 3. Wicked problems do not have solutions that true-or-false, but good-or-bad | Tame problems have identifiable solutions. The problem solver knows whether the solution is right or wrong, true or false. | There are no true or false answers. Solutions to wicked problems impact many parties and these impacts may vary. Some may perceive the impacts to be desirable and good while others may find the impacts are harmful and undesirable | <u>Siting of a nuclear plant</u> When a nuclear plant is sited, some people in a nearby community may be strongly opposed to it and have concerns about safety, whereas others may be pleased about the plant site and possible economic opportunities the siting may bring |
| 4. Do not have an immediate or ultimate test of a solution | It is possible to immediately determine whether a solution attempt has been successful. The test of a solution is under the control of a limited number of people. For example, designing the interior of a house. | Solutions to wicked problems generate "waves of consequences" over an extended period of time. The solution attempt cannot be fully evaluated until the waves of consequences have been allowed to play out and are themselves assessed. | <u>Nuclear safety</u> A probabilistic risk assessment (PRA) may show a reactor to be 'safe' but that does not mean that an accident can never occur |
| | | Planners typically satisfice and stop studying outcomes due to resource constraints. | |
| 5. Every solution attempt has real world consequences | Many problem solutions may be attempted analytically, in a simulation or in a lab with little to no impact on the real world. Because of this minimal impact, a very large number of solution attempts are possible. Trial and error is an acceptable approach to problem-solving. | Every implemented solution has real-world consequences, creating traces that cannot be undone; "half-lives" of consequences are very long. Every trial counts. | Nuclear innovation<br>Decisions to fund a particular nuclear technology or design, also involve decisions to not fund others. These unfunded designs and companies may not continue and may be shut down. |
| 6. Wicked problems do not have a finite set of potential solutions | There is a well-defined solution or a set of possible solutions. | There may be a vast number of unknowable solutions or even the absence of a single solution. The size of the solution set cannot be quantified because the number and "goodness" of the solutions depend on the entities impacted by the problem and solution. | Nuclear waste<br>How should a nuclear waste repository be designed to minimize societal and environmental impact? Who is considered while evaluating impact? If the set of people, communities, and non-human actors considered expands, the possible facility design solutions also expand |
| 7. Each wicked problem is unique | These problems may be arranged into 'classes' or 'types' of problems. | While wicked problems may share some similarities, there are likely to be additional distinguishing properties that make it impossible to replicate the solution of a wicked problem attempted elsewhere. | Nuclear Waste<br>Approaches to nuclear waste siting which have been tried (successfully) in Scandinavian countries may not work if replicated exactly in the US because of the unique history of the nuclear sector and its impacts on communities |
| 8. Every wicked problem is a symptom of another wicked problem | These problems may be connected to each other but not necessarily or always. | Wicked problems are symptoms of other problems. One must carefully choose the 'level' at which the problem is solved and choose a level that is neither to high and 'abstract' nor too specific and concrete | Nuclear proliferation<br>Nuclear proliferation is the symptom of another 'problem' - the discovery of nuclear fission and the spread of nuclear technologies around the world. Nuclear proliferation as a problem could be solved by reversing the spread of nuclear technology (not possible |

| | | | but attempted in a limited sense) or policing every instance of use of nuclear technology (attempted but also in a limited sense) |
| --- | --- | --- | --- |
| 9. A wicked problem can be represented or framed in more than one way | These problems can typically be represented simply and without discrepancies | Representation and explanation of wicked problems depend upon the person explaining or representing the problem. Problem solvers tend to pick the representations that best fit the solutions that are available to them. | Nuclear safety<br>A nuclear plant designer may view safety as a design problem, a regulator might view safety as a problem of sufficient oversight of plant designers and operators, and a plant operator might view safety as being a problem related to organizational pressures and insufficiency of resources to ensure safe operation |
| 10. The planner or policymaker has no right to be wrong | Trial and error is an acceptable solution approach as solution attempts have little real-world impact. | Planners are morally responsible for every solution attempt which has real-world impacts that might be far-reaching. | Cost of electricity<br>Failures to design a plant carefully or manage a construction project well may lead to significant cost escalations which could increase electricity prices for hundreds of thousands of households |

An elective course

The course is an elective with the students in the class last year (Winter 2023) being a mix of senior undergraduate students from the Department of Nuclear Engineering and Radiological Sciences (NERS) as well as masters and Ph.D. students from the department. While the majority of students enrolled in the course this term are again from NERS, the course also includes students from Chemical Engineering, Computer Science, Mechanical Engineering, and the Business School. On the first day of class when students are invited to share what brought them to the course, the students from outside NERS shared that their interest in nuclear energy and technologies and in some cases, internships or approaching employment in the nuclear sector drew them to the course. In the first year of the course, the course was only advertised internally within the department, leading to an enrollment of 10 students. The number of students enrolled in the course has doubled this year (with some students deciding whether they will take the course for credit or audit it). Students auditing the course are not required to complete weekly assignments or the term project but are encouraged to participate in the "Nuclear in the News" exercise and complete the required readings before each course session.

It should be added that while the course is strictly an elective for most students, doctoral students in NERS taking the fission-policy track are required to take the course and are tested on it as part of their written candidacy exam.

Course structure

The course unfolds over sixteen weeks with two eighty-minute lectures a week. The list of topics covered in the lecture is shown below. Not included in the list of topics are the sessions devoted to student presentations. Following the initial lecture on the framework of wicked problems and how this framework can be applied to problems examined as part of this course, students are led through sessions on social construction of technologies and a history of nuclear technologies, sessions on innovation, technology selection and the role of startups, and global transfers of nuclear technologies. Following these initial sessions on history and innovation, the course pivots for a week to focus on economics and financing. A session on speculative design is included in the Winter 2024 course offering as a resource for the students for their term projects. Following this, a series of sessions focus on safety, risk, regulation, and accidents. The course then shifts to focus on the siting of nuclear facilities. This includes sessions on nuclear waste, consent based siting, siting of new nuclear facilities, and the impacts of legacy facilities. These sessions on siting are followed by two lectures on security and non-proliferation. A final session on reimagining nuclear engineering – which invites the students to envision possible futures for the nuclear field, including futures in which wicked problems have been solved – wraps up the lectures. The semester itself ends with sessions in which students present their projects.

**Table 2** A list of lecture topics are included below.

| Lecture topics |
|---|
| Introduction to the course + Nuclear policy problems as wicked problems |
| Technologies as socially constructed + a brief history of the nuclear technologies and institutions |
| The role for nuclear energy in a clean energy system? |
| Innovation in the nuclear sector |
| Technology selection, demonstration, and the role of startups |
| Technology transfer – The nuclear industry as a global industry |
| Cost - a primer on engineering economics |
| Financing nuclear projects |
| Speculative design |
| Safety – what is a safe reactor? |
| Risk – The engineer's framing of risk |
| Risk – Framing, perception, and reality |
| Accidents and what we can learn from them |
| Nuclear waste: a history of management and mismanagement |
| Consent-based siting and nuclear waste management |
| Siting nuclear facilities |
| New Nuclear Communities |
| Legacy, contamination, and cultural heritage |
| Uranium mining and markets: local and global impacts |
| Security – security by and for whom? |
| Nuclear energy, equity, environmental justice |
| Reimagining nuclear engineering - what does a reflexive, inclusive field look like? |

Each lecture is informed by a series of guiding questions the students are expected to be able to knowledgeably reflect and write about based on their learnings from each lecture. For example, the fourth and fifth course sessions focus on innovation in the nuclear sector and technology selection, demonstration, and the role of startups respectively. The guiding questions for these two sessions are shown in the Table 3 below.

**Table 3.** Guiding questions for the sessions on innovation in the nuclear nuclear sector and technology selection, demonstration, and the role of startups respectively

| Innovation in the nuclear sector | Technology selection, demonstration, and the role of startups |
|---|---|
| What do we mean by 'innovation'? | The technology lifecycle and the emergence of dominant technologies |
| Where, when, and how does innovation take place? | Should the state pick 'winners' and shape the emergence of winning technologies? |
| What is the role of the state in stimulating or stewarding innovation? | Are the winning technologies always the 'best' technologies? |
| Where does innovation occur in the nuclear sector? | How did the light water reactor become the dominant technology? |
| Has innovation in the nuclear sector been equitable historically? Can innovation in the nuclear sector be equitable in the future? | Are we likely to see the emergence of dominant reactor technologies in the future? |
| The importance of innovation and maintenance | What role do large companies vs. startups play in an innovation ecosystem? |
| | How does private vs. government funding impact innovation decisions? |
| | What is the state of innovation and technology development in the fusion sector? |

Students are asked to complete (typically) two to three readings before each lecture. For students interested in exploring the topic further, a number of additional optional readings are also offered. Students in the initial course offering in Winter 2023 found these readings especially helpful particularly when their term projects were on a related subject. As an additional resource, students are given an annotated bibliography template and encouraged to fill this out as they complete their readings over the course of the semester. The annotated bibliography is not graded. Students are encouraged to create it so that it might serve as a resource for them to be able to revisit the course readings after the completion of the semester.

<u>Nuclear in the news</u>

Class participation and discussion is central to the course and students are encouraged to ask questions throughout the lecture (and do) starting from the very first course session. Beyond the active discussion encouraged in every lecture students, another significant opportunity for class participation is at the start of every lecture. The first fifteen to twenty minutes of each class are devoted to discussing nuclear technology topics in the news (this the 'Nuclear in the News') component of the course. At the start of the course, the students are invited to select two topical keywords (a list of the keywords is shown in Table 4 below) and read a total of two news articles before each session (or peruse other sources of information — such as podcasts, documentaries, or youtube videos). As a starting point, students are offered an initial set of potential resources and news outlets where they may begin their reading but are encouraged to read broadly and identify new sources and share them with the class. When perusing these different sources of information, students are encouraged to read between the lines and reflect critically on the source of the information, the affiliation of the author, potential biases of the author or publication outlet, and how these biases might color what is being written about. Prior to each lecture session, students summarize both what they have learned as well as reflect critically on the article or other source on sticky notes on an online course mural shown in Figure 1 below. Additionally, students are encouraged to find connections (marked as connecting lines on the mural) between their readings and those of their classmates or between current and previous readings and identify trends or connections where they may exist.

**Table 4.** Nuclear in the News keywords

| Nuclear in the News keywords | |
|---|---|
| Decarbonization | Nuclear construction |
| Coal to nuclear | Critique of the nuclear sector |
| Nuclear supply chain | Nuclear regulation |
| Reprocessing Technology | Microreactors |
| Spent fuel/waste management | Decommissioning |
| Uranium enrichment | Government |
| Uranium mining and markets | Nuclear safety |
| Small modular reactors (SMRs) | Nuclear security |
| Demonstration projects | Fusion Technology |
| Contamination and cleanup | Nuclear Innovation |
| Space propulsion | |

**Figure 1.** The figure below shows the "Nuclear in the News" Mural

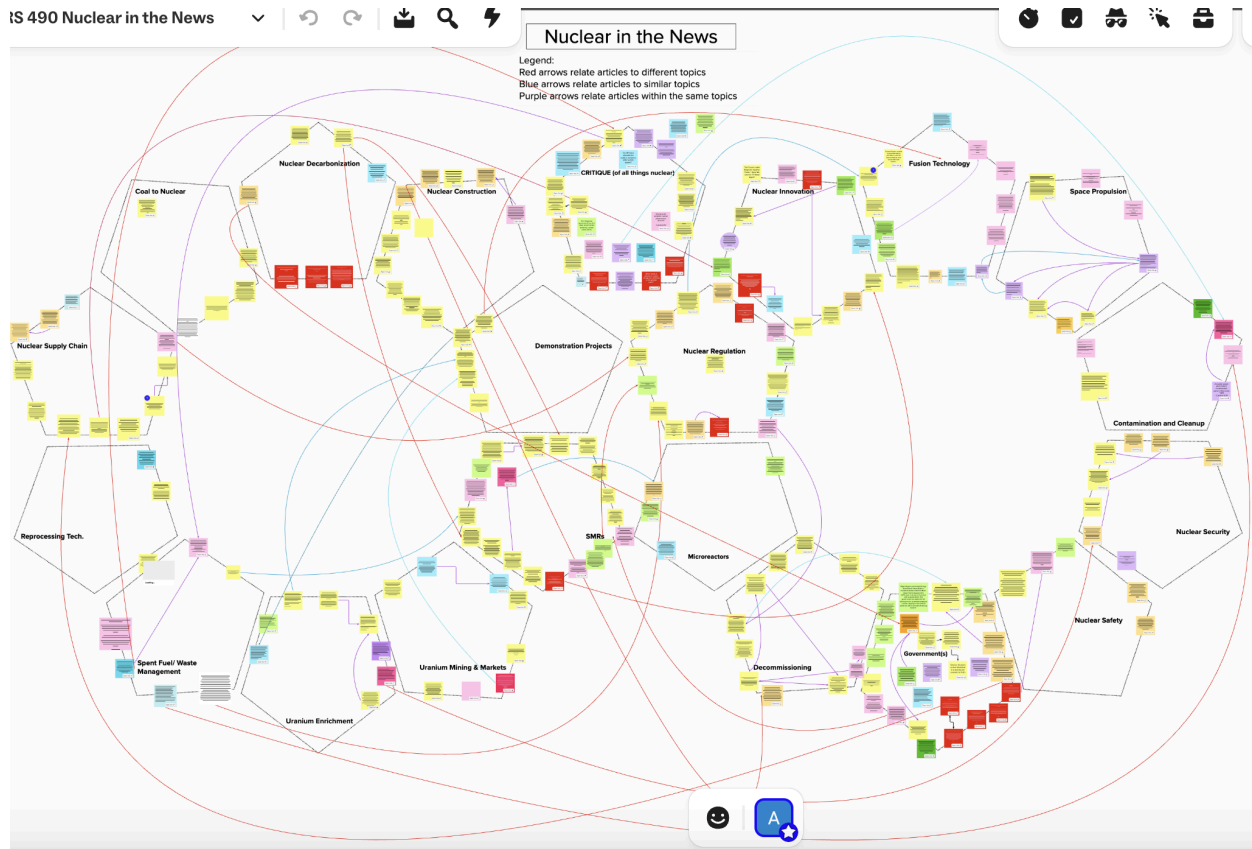

Use of technology in the classroom

In addition to Mural – which is used for the Nuclear in the News segment, as well as for in-class workshops, the course also makes use of Menti. Menti is a helpful tool for eliciting student responses live during a lecture. Gathering student responses through Menti is particularly helpful for gathering background information or understanding of a particular problem or topic by the students. For example, Figure 2 below shows student responses gathered using Menti to the question "what does the word 'policy' mean to you?". Students were asked this question on the first day of class.

**Figure 2.** Student responses on the first day of class when asked "What does the word 'policy' mean to you?"

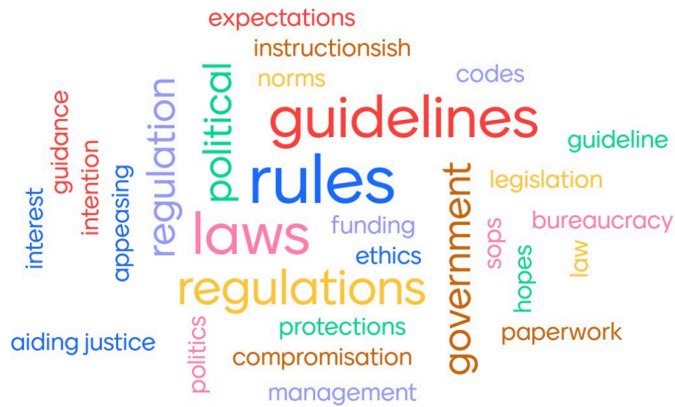

Course readings and resources

The course does not have a required textbook. As noted above, students are required to complete two to three readings prior to each course session. Additionally, the syllabus includes as a resource several readings on the historical, sociological, and anthropological studies of nuclear organizations and institutions, as well as a list of books on the history and sociology of technology broadly. The syllabus also lists several podcasts as resources. These include "Press the Button", "The Sketch Model Podcast", "The Received Wisdom", and "Things that go Boom".

In addition to these resources, all students in the course in Winter 2024 received a copy each of two books on speculative design – The Extrapolation Factory[48] and Speculative Everything[49]. These books offer conceptual frameworks and methods students are encouraged but not necessarily required to explore as part of their term projects.

Suffusing questions of ethics and justice across sessions

While the course is not solely about questions of ethics and justice – these themes suffuse every session of the course. For example, in the initial sessions on innovation, students are invited to think critically about innovation practices: is innovation necessarily always desirable and ethical? Who has a voice in the innovation process (and who doesn't)? How, in the process of innovation and design, can inequities become embedded in our technologies? What does it mean for companies, researchers, and government agencies to make ethical and judicious use of public funding?

In sessions on safety and risk students reflect on how experts and publics (plural as the public is not a monolithic group) differently perceive and frame risks and how engineers, instead of dismissing public perceptions of risk as affective and irrational, can make sense of them and try to engage meaningfully with publics, without 'acceptance' of a technology being an instrumental goal framing such engagement. In the sessions on nuclear waste management, nuclear facility siting, legacy, contamination, and cultural heritage, students reckon with the complicated history of the nuclear sector and often disproportionate impacts of nuclear technology development (weapons technologies especially) on indigenous communities in particular. They reflect on what measures might be taken to correct historical wrongdoings as well as what measures might be taken to ensure responsible and ethical development and use of nuclear technologies in the future – and how communities might be able to directly participate in this decision-making.

As a culmination of all of this thinking, the final course session before the student presentations is on "Reimagining nuclear engineering - what does a reflexive, inclusive field look like?". In this course session, the students discuss and reflect on what a socially and environmentally responsible, reflexive, and inclusive nuclear field could look like and what institutional structures, engineering practices, and power dynamics need to change to achieve this shift. As an assignment concurrent to this session students are invited to do three things:

(1) Write a code of ethics for nuclear engineers informed by all that they have learned throughout the course
(2) Draft a letter to their future selves describing what brought them to this course, what 'wicked' problems in the nuclear sector they hope their future selves will have helped solve, and what they have learned over the course of the semester that they would like your future self to remember. Students are encouraged to use a service such as futureme.org to send this letter to their future selves.
(3) Envision and visualize possible futures for nuclear technologies as well as nuclear professionals in the year 2100. Some questions students asked to consider while imagining these futures are shown in Table 5 below. Students are asked to use AI image generators to create images depicting these futures.

**Table 5.** Prompts for envisioning nuclear futures

| Prompts for envisioning nuclear futures |
| --- |
| What do nuclear energy technologies - fission and fusion – look like in 2100 and how and where are they being used? Who is using these technologies? |
| Have we averted the worst effects of climate change? If so, what role have nuclear energy technologies played in accomplishing this? |
| How much nuclear waste have we produced? Where is that nuclear waste being stored? Is that waste still regarded as 'waste'? |
| Have we repaired the legacy environmental and health impacts of nuclear technologies? |
| How many nuclear weapons exist in 2100? Who has these weapons? If there are no weapons, what have we done to prevent countries from rearming? |
| How are nuclear materials safeguarded in 2100? |
| What roles do nuclear engineers play in society? Are we trusted? |
| What are some new cutting-edge areas of nuclear science and technology that have emerged? (quantum computing, space nuclear propulsion, transformative new materials for use in extreme environments?) |

**Course deliverables**

In addition to periodic assignments are required to complete two key deliverables – an op-ed and a term project.

Periodic Assignments

Students complete periodic assignments. These assignments do not have a weekly cadence. For the first course offering, the students completed four such assignments. As part of the second course offering the number of assignments has been increased to six. The timing and cadence of the assignments are aligned with other recurring deliverables (such as the op-ed, mid-semester and final presentation, and the final term paper). These periodic assignments, either call on students to delve deeper into a reading, to reflect on it, or to collect and analyze original data. For example, in the very first course assignment, students choose a problem at the intersection of technology, policy, and society and then, using the Wicked Problems paper as a framework, examine whether the problem they have chosen meets each of the criteria for wicked problems as identified in the paper, and whether the problem they have chosen has other aspects of wickedness not captured in the paper. In a subsequent paper, having completed readings on technological lock-in and dominant designs (and having discussed these topics and theories in class), students are asked to imagine that they are in a position of leadership at the DOE and tasked with designing a program of research and innovation funding that meets the following criteria: (1) pursuit and development of if not the 'best' then at least clearly superior technology options and designs; (2) A moderately rapid pace of innovation and technology commercialization, attuned to the urgencies of the looming climate crisis; (3) responsible and accountable use of taxpayer dollars. Another example of an assignment is one in which students collect and analyze the mission and vision statements of nuclear companies (fission and fusion), along with details on the source of funding of the companies and the backgrounds of their founders. Students are then asked to analyze where there are patterns that appear across and within the fission and fusion companies, and what their data tells them about the nature and state of innovation in the nuclear sector, as well as what and who is driving it.

## Op-ed

The inclusion of the op-ed as a course deliverable was a deliberate decision stemming from the instructor's desire that engineers ought to be able to communicate clearly and effectively with publics and well as decision-makers. Students in the course are invited to author an op-ed on any topic of their choosing (selected in consultation with the instructor) and are welcome to submit the op-ed at any point in the semester. Many students chose op-ed topics that were closely related to their term projects. The only significant limiting criterion for the op-ed was the word limit (a range of 600 to 1000 words). Many students expressed that while they found it challenging to express their viewpoints concisely, they appreciated the challenge and the opportunity to develop a new set of skills through the writing of the op-ed. Op-ed topics from the first course offering included the impact of the Russian-Ukrainian conflict on the commercial nuclear industry, the need for nuclear energy in a low-carbon economy, the environmental and societal challenges associated with mining uranium, and the need for the US to adopt a no first use nuclear weapons policy, to name some. In the Winter 2024 course offering, selected Op-Eds will be invited for publication as part of a new ANS NSTOR open access collection on "Reimagining Nuclear Futures: Emerging Voices on Technology, Policy, and Society".[50]

## Term project

In addition to periodic assignments, and an op-ed, each student in the course also completes a term project. Students select and define their term project (in consultation with the instructor) by the mid-semester mark and present their mid-semester progress to the class. The term project deliverables include a written research paper (five to seven thousand words) as well as a final presentation. During the first course offering, all student term projects were research-based. Students were offered a range of possible term paper topics by the instructor while also being given the option of designing a topic of their choosing. From the first course offering, student term projects include:

1. The value of flexible, load-following reactors in distributed nuclear-renewable energy systems
2. Expert assessments of the prospects for commercial fusion energy
3. Regulatory and environmental challenges and opportunities for nuclear fusion
4. Innovative financing approaches for SMRs
5. Stockpile modernization across nuclear weapon states

For the second-course offering in Winter 2024, the term projects are a creative rather than research endeavor. Students have been tasked with interpreting, exploring, or offering solutions to nuclear policy problems through the lens of storytelling or play. While student projects are still at an early stage – it is clear that students are interested in exploring a range of mediums through their work including music, documentary videos, podcasts, and video games.

Tutorials and Workshops

In addition to lectures, students also participate in workshops and tutorials that take place during the regular class sessions. During the first course offering students received a tutorial on how to write op-eds as well as a tutorial on how to use a GIS-based nuclear energy facility siting tool developed at the University of Michigan at the Fastest Path to Zero.

Examples of workshops include one on the nuclear innovation ecosystem. As part of this workshop, students, working in teams of two to three, were tasked with reflecting on the purpose of innovation in the nuclear sector, identifying key actors (individuals, organizations, government agencies), and marking how these actors are connected by flows of funds, people, and information. This exercise quickly revealed to the students the complex, interconnected, convoluted, and not perfectly knowable nature of an innovation ecosystem. Through this exercise, they learned to appreciate that a key feature of these ecosystems is the asymmetry of information as held by various actors, and the critical role that connecting organizations – including government agencies can play in stewarding and tending to these systems of innovation. These observations resonate strongly with research on the role of governments – as clearinghouses of innovation[51]– in modern industrial ecosystems. (Students learned about this theory in a subsequent course session).

Grading

The final grade students receive for the course is distributed in such a way that no single deliverable is overwhelmingly significant. This choice was made deliberately for two reasons:

(1) the most beneficial learning experience in this course is one in which students are actively engaged in every aspect of the course (readings, in-class discussion, periodic assignments, op-ed development, and term project) throughout the semester as each element of the course supports the students in achieving a deeper and more layered understanding of the course material.

(2) Given that the course is non-traditional and calls for the development and exercise of what for most students will be a new set of skills and ways of thinking, overweighting any particular assignment or deliverable is likely to create an unnecessarily stressful, high-stakes which is ultimately counterproductive to learning in this course. As a result, no single element of the course counts significantly, because all elements are important. This is among the first pieces of information students receive at the start of the semester.

Course outcomes

In its first offering, the course received overwhelmingly positive student evaluations. In their course evaluations students, when asked to comment on the quality of the course students offered unanimously positive input. They commented on (paraphrased or quoted verbatim from student responses):

- The "exceptional" quality of the course
- The "robustness" of the syllabus
- The wide array of topics and an appreciation for the breadth of material covered
- An appreciation that the topics covered applied not only to nuclear engineering but "engineering and science in general"
- Thoroughness of the materials covered
- Having learned a lot of valuable things they could use in their career
- The engaging nature of the lectures and discussions
- The instructor's willingness to adapt the flow of the lecture to the interests of the students
- The instructor's respectful treatment of the students in the course

When asked for areas of improvement for future course offerings, students suggested that they would have appreciated more time to complete readings, which could be posted earlier, possibly more than a week ahead of each lecture, as well as better pacing course deliverables towards the end of the semester when several deliverables were due in quick succession. These recommendations have been incorporated into the current course offering.

To the statement "I think it is important for nuclear engineers to learn about the social, policy, and ethical implications of nuclear technologies" all students chose the "strongly agree" option. Similarly, all students also chose the "strongly agree" option in their course evaluations that all nuclear engineers should take the course.

**Discussion: Recommendations for those developing similar courses**

The initial course offering and the second course iteration currently in progress, as well as the student evaluations suggest that the course development efforts have been successful. While there are specific aspects of lectures, including readings, and the details of assignments, that will continue to evolve over the years, it is the instructor's intent that the course structure described here will be sustained.

For those who are considering developing and offering similar courses at their own institutions, the course framings, topics, and structure – including readings, deliverables, in-class workshops and tutorials as they are implemented in the course – might offer a starting point for course development. Alternatively, those who wish to replicate the course syllabus are welcome to do so. A few aspects of the course that make it a meaningful learning and teaching experience bear repeating:

(1) As is evident from the wicked problems framing, and as explicitly stated, problems at the intersection of technology, policy, and society do not yield to well-defined and permanent solutions. Often these problems and their potential solutions are viewed in diametrically opposite ways by different entities. This is a fundamental feature of these problems that must be acknowledged. Doing so can lead to intellectual discomfort – particularly as the sciences and engineering – built as they are around positivism and seeking the 'best' solutions. However, as noted earlier, wicked problems do not have 'best' solutions – they only have good or bad ones such that goodness and badness lie in the eye of the beholder or assessor. Therefore, in a course such as this one, it is important for students to hold opposing truths, be comfortable with not being able to find 'right' answers but also appreciate that these are not reasons for not looking for *any* answers, but instead, for trying harder and looking further.

(2) A second key aspect of this course, as described earlier, is that it is important for students to be able to both learn how to think and what to think about wicked problems. This is a crucial skill to develop as the manifestation of these problems will continue to evolve over time.

(3) A third important aspect of the course relates to viewing the classroom – both students and the instructor – as a community of learning and practice. The course is as much an opportunity for the instructor to learn about and question their own assumptions about problems at the intersection of technology, policy, and society (assumptions that come to light while responding to student questions – which themselves may be founded in still other assumptions). For this reason, building a teaching and learning environment in which students feel comfortable asking questions, even offering opposing viewpoints, is essential. In this manner the classroom becomes a living laboratory[52] for the exploration of wicked problems.

**Figure 3.** A transformation in nuclear engineering from a discipline that narrowly views our roles as nuclear engineers and the interactions of our technologies with society to a richer, more complex (and complicated) understanding of the origins and impacts of our technologies.

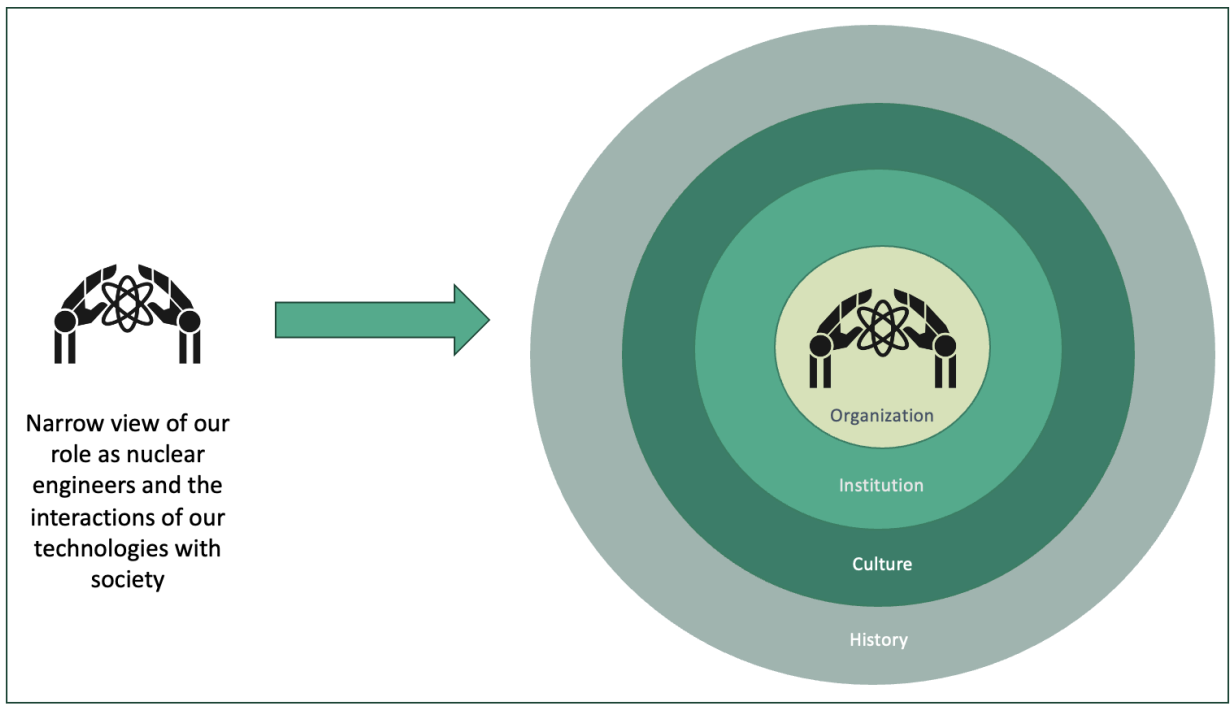

Narrow view of our role as nuclear engineers and the interactions of our technologies with society

Organization
Institution
Culture
History

## Conclusions

This paper has described the design, implementation, and outcomes of a course on Nuclear Technology, Policy, and Society at the University of Michigan. As noted at the start of this paper, topics covered in this course have historically been regarded as 'soft', 'non-technical' and not relevant to an engineering education. However, the design and implementation of this course is proof that topics and problems at the intersection of technology, policy, and society can and should be treated in rigorous ways – even if these problems do not yield everlasting solutions, and even if these solutions are not universally accepted or satisfying. The course outcomes – in particular the student evaluations, suggest that students find the course to be a valuable, even essential part, of their nuclear engineering education. All of this suggests a potential repositioning of how we view what constitutes nuclear engineering: We have typically understood our technologies and conceptualized our role as nuclear engineers in narrow ways. We have concerned ourselves with the science and engineering details of our technologies and designs. Our conceptualization of what constitutes engineering and our conceptualization of our roles as engineers in society can be broader, particularly as the field matures, as it is now doing. As part of this broader view, we should think of our technologies and our roles relative to them in the fullness, richness and complexity of how our technologies interact with society, and in so doing, we must expand our notion of what it means to be a nuclear engineer.